\documentclass[onecolumn,secnumarabic,amssymb, nobibnotes, aps, pr,superscriptaddress]{revtex4-1}

\def\l{\left}
\def\r{\right}

\newcommand{\f}{\frac}

\newcommand{\mean}[1]{ \l<#1\r> }

\usepackage{amsmath,amssymb}
\usepackage{collectbox}
\usepackage{stackengine,graphicx}
\usepackage{bbm}
\usepackage{capt-of}

\begin{document}

\title{Assimilated LVEF: A Bayesian technique combining human intuition with machine measurement for sharper estimates of left ventricular ejection fraction and stronger association with outcomes.}


\author{Thomas~McAndrew, Ph.D}
\email{tmcandrew@crf.org}
\affiliation{Cardiovascular Research Foundation, 1700 Broadway, New York , NY , USA}

\author{Bjorn~Redfors, M.D. Ph.D}
\affiliation{Cardiovascular Research Foundation, 1700 Broadway, New York , NY , USA}

\author{Aaron~Crowley, M.A.}
\affiliation{Cardiovascular Research Foundation, 1700 Broadway, New York , NY , USA}

\author{Yiran~Zhang, M.S.}
\affiliation{Cardiovascular Research Foundation, 1700 Broadway, New York , NY , USA}

\author{Maria~Alu, M.S.}
\affiliation{Cardiovascular Research Foundation, 1700 Broadway, New York , NY , USA}
\affiliation{Department of Medicine, Division of Cardiology , Columbia University Medical Center/New York-Presbyterian Hospital , New York , NY , USA}

\author{Matthew~Finn, M.D.}
\affiliation{Cardiovascular Research Foundation, 1700 Broadway, New York , NY , USA}
\affiliation{Department of Medicine, Division of Cardiology , Columbia University Medical Center/New York-Presbyterian Hospital , New York , NY , USA}

\author{Ariel~Furer, M.D.}
\affiliation{Cardiovascular Research Foundation, 1700 Broadway, New York , NY , USA}
\affiliation{Department of Cardiology, Tel-Aviv Medical Center, Tel-Aviv, Israel}
\affiliation{Internal Medicine 'T', Sourasky Medical Center, Israel}

\author{Shmuel~Chen, M.D., Ph.D.}
\affiliation{Cardiovascular Research Foundation, 1700 Broadway, New York , NY , USA}

\author{Geraldine~Ong, M.D.}
\affiliation{Cardiovascular Research Foundation, 1700 Broadway, New York , NY , USA}
\affiliation{Division of Cardiology, St. Michael's Hospital, 30 Bond St, Toronto, ON M5B 1W8, Canada.}

\author{Dan~Burkhoff, M.D.}
\affiliation{Cardiovascular Research Foundation, 1700 Broadway, New York , NY , USA}
\affiliation{Department of Medicine, Division of Cardiology , Columbia University Medical Center/New York-Presbyterian Hospital , New York , NY , USA}

\author{Ori~Ben-Yehuda, M.D.}
\affiliation{Cardiovascular Research Foundation, 1700 Broadway, New York , NY , USA}

\author{Wael A. Jaber, M.D.}
\affiliation{The Cleveland Clinic Foundation (M.Y.), Cleveland, Ohio;}

\author{Rebecca~Hahn, M.D.}
\affiliation{Department of Medicine, Division of Cardiology , Columbia University Medical Center/New York-Presbyterian Hospital , New York , NY , USA}

\author{Martin~Leon, M.D.}
\affiliation{Cardiovascular Research Foundation, 1700 Broadway, New York , NY , USA}
\affiliation{Department of Medicine, Division of Cardiology , Columbia University Medical Center/New York-Presbyterian Hospital , New York , NY , USA}

\date{2018-05-07}



\begin{abstract}
The cardiologist's main tool for measuring systolic heart failure is left ventricular ejection fraction (LVEF).
    Trained cardiologist's report both a visual and machine-guided measurement of LVEF, but only use this machine-guided measurement in analysis.
 We use a Bayesian technique to combine visual and machine-guided estimates from the PARTNER-IIA Trial, a cohort of patients with aortic stenosis at moderate risk treated with bioprosthetic aortic valves, and find our combined estimate reduces measurement errors and improves the association between LVEF and a 1-year composite endpoint.
\end{abstract}

\maketitle
\section{Introduction}

\label{intro}

Heart failure (HF)~\cite{hunt20092009,mcgrady2013n,leon2010transcatheter,leon2016transcatheter,ho1993epidemiology} remains a leading cause of patient mortality~\cite{ho1993survival,consensus1987effects,cleland2005effect} and reduced quality of life~\cite{witte2005effect,jaarsma2000self,demers2001reliability}.
The scientific communities' link between declining left ventricular ejection fraction (LVEF) and a higher probability of adverse events~\cite{gavazzi1997value,christenson1995effect} supports the need for more accurate and reliable methods for measuring LVEF.

While Simpson's method for measuring LVEF is more accurate than visual estimation, a cardiologist's visual estimate can reduce LVEF variability across patients (increasing reproducibility) by drawing on past experience and empirical self-correction~\cite{gudmundsson2005visually,bellenger2000comparison,otterstad1997accuracy}.
Previous studies found Simpson's biplane method reproducible across diverse types of patients measured by the same echocardiographer, but different echocardiographers and different echocardiographic tools can draw different conclusions from the same patient~\cite{bellenger2000comparison,otterstad1997accuracy}.
Identifying this extra variability in LVEF measurements can create more robust LVEF estimates and may translate to better patient diagnosis, treatment options, and prognostic predictions.

We applied a simple Bayesian approach to fuse the more stable cardiologist's visual LVEF with the more accurate Simpson's bi-plane calculated LVEF.
This novel assimilated LVEF allows expert opinion and clinical experience to more directly take part in analysis, reduces the measurement variability in LVEF, and leads to more robust inference between LVEF and outcomes.

\section{Methods}
\label{methods}

We studied echocardiographic data collected from the PARTNER-IIA trial~\cite{leon2016transcatheter,leon2010transcatheter} which randomized North-American patients with severe aortic stenosis to either surgical aortic valve replacement (SAVR) or transcatheter aortic valve replacement (TAVR).

An independent core-lab measured LVEF in $1366$ PARTNER-IIA patients using both visual estimation and Simpson's bi-plane method.
The visual estimation of LVEF was specified in commonly used intervals of $5\%$ and Simpson's biplane method was reported to the first decimal place.

We expressed our uncertainty in a patient's true LVEF by placing a normal distribution on top of the cardiologist's visual LVEF estimate with a 18.1\% standard deviation reported from previous work~\cite{van1996comparison}.
The next step overlaid a normal distribution over top of the cardiologist's measurement of LVEF with Simpson's biplane method with a $8.8\%$ standard deviation reported from previous work~\cite{chuang2000importance}.
Our final step combined both visual and Simpson's biplane normal distributions with Bayes theorem.

Mathematically, we approximated a patient's true LVEF $(\theta)$ by combining the visual estimate $(V)$ with variance $\sigma[V]$ and Simpson's biplane estimate $(S)$ with variance $\sigma[S]$ as
\begin{equation}
  p(\theta = l | \mathcal{D}, \sigma ) \propto p(S | \theta,\sigma \l[S\r]) \times p(\theta | V,\sigma\l[V\r])
\end{equation}

where $\mathcal{D}$ represents the visual LVEF and Simpson's LVEF data and $\sigma$ represents the measurement error (or uncertainty) in both LVEF measurements.
We updated our visually informed prior, $p(\theta | V, \sigma\l[V\r])$, with our uncertainty in Simpson's biplane measurement, $p(S | \theta, \sigma\l[S\r])$, and expressed our final uncertainty in a patient's true LVEF as 
\begin{equation}
  p(\theta | \mathcal{D}, \sigma) \propto \mathcal{N}\l( \theta,\sigma[\theta] \r)
\end{equation}
where $\mathcal{N}$ represents a Normal distribution with $\theta$ equal to
\begin{equation}
   \theta = \dfrac{\sigma[V] S + \sigma[S] V}{ \sigma[V] + \sigma[S]  } 
\end{equation}
and $\sigma[\theta]$ equal to
\begin{equation}
  \sigma[\theta] = \dfrac{1}{ \sigma^{-1}[V] + \sigma^{-1}[S]}.
\end{equation}

\begin{figure}[!p]
  \centering\includegraphics{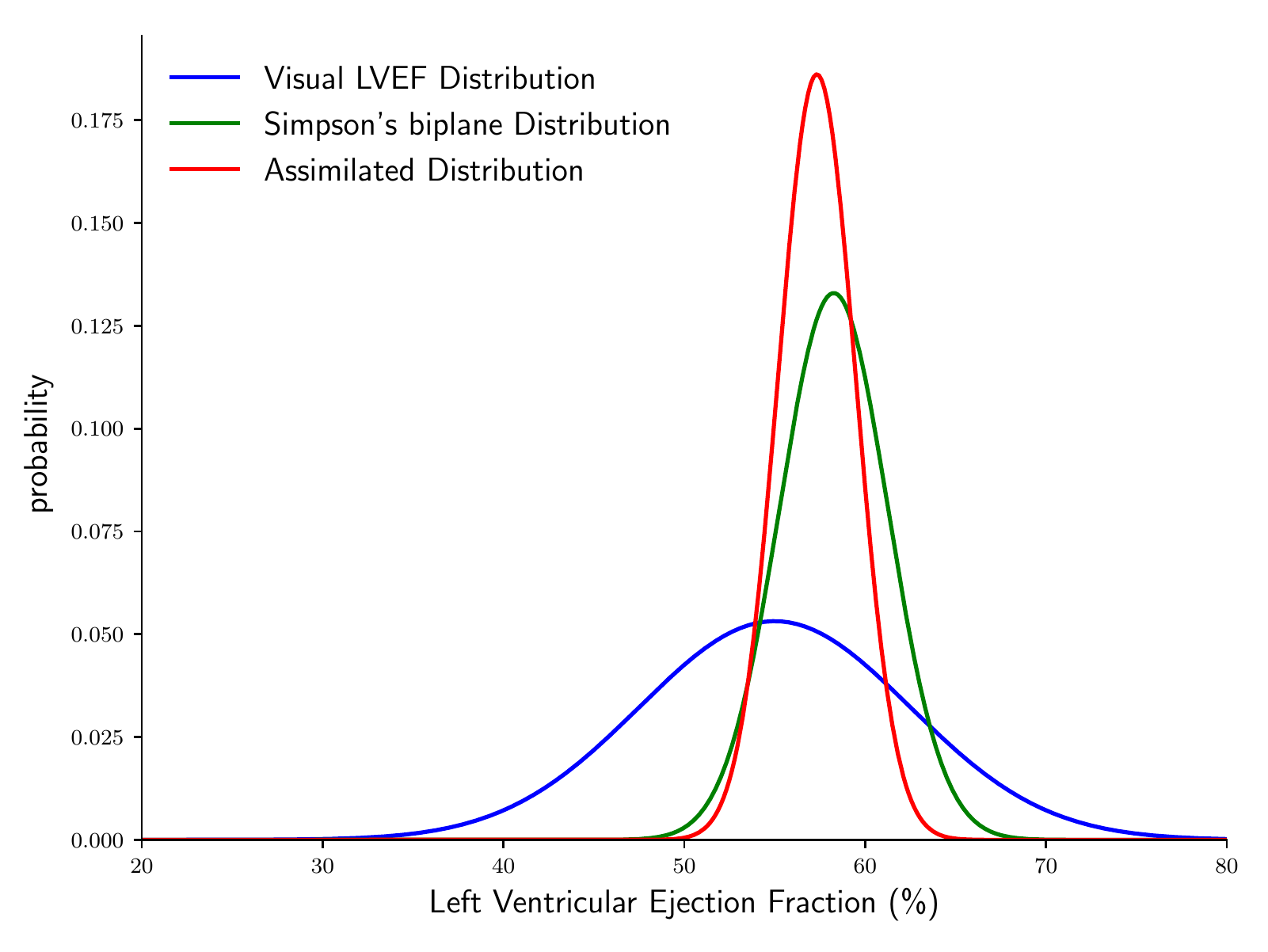}
  \caption{
    We combine a patient's visual LVEF and Simpson's biplane LVEF measurements for a more assured estimate of a patient's true LVEF. 
    This assimilated LVEF measurement draws accuracy from Simpson’s biplane LVEF and consistency from the visual LVEF to reduce measurement error. 
    \label{fig1.assimExample} }
\end{figure}

We can redefine $\theta$ by dividing Simpson's biplane precision by the visual precision 
\begin{equation}
  \omega = \f{\sigma^{-1}[S]}{\sigma^{-1}[V]}
\end{equation}
and total variation
\begin{equation}
  T = \sigma[V] + \sigma[S].
\end{equation}
After rewriting $\theta$ and $\sigma[\theta]$ in terms of $\omega$ and $T$, we find $\theta$ borrows Simpson’s biplane accuracy and lies $\omega$ times closer to Simpson’s biplane measurement than the visual estimate,
\begin{equation}
  \theta_{\text{MAP}} = \f{\omega S + V}{T}
\end{equation}
and used the cardiologist's visual estimate (their experience) to reduce measurement error by
\begin{equation}
  R = \f{\sigma[\theta] - \sigma[S]}{\sigma[S]} = \f{-1}{\omega + 1}
\end{equation}
relative to Simpson's biplane measurement (Figure~\ref{fig1.assimExample}). 

We can build uncertainty around the 18.1\% visual LVEF's measurement error and 8.8\% Simpson's measurement error to quantify how well this assimilated method reduces measurement error $(R)$.
For each measurement error, we consider a non-informative Gamma-distributed prior and Gamma likelihood.
We draw $5 \times 10^{3}$ parameter samples from a $2 \times 10^{4}$ long Markov-chain Monte-Carlo chain with Normally distributed proposal for both visual and Simpson's measurement error, and from each parameter sample, draw a predictive measurement error.
Each pair of (visual, Simpson) measurement error predictions corresponds to a relative reduction and can quantify our uncertainty in $R$.

We also applied the above measurement error model to create more robust statistical inferences in time to event analysis.
Our first step draws samples from each patient's LVEF distribution.
This first step represents the measurement-error step of our model.
The next step applies Kaplan-Meier methodology and Cox-proportional hazards model to predict $1$ year death, rehospitalization, and stroke as if each LVEF measurement was exact.
The last step compiles the N test statistics created from the novel assimilated LVEF’s predictions to conclude how differences in LVEF affect the $1$-year event rate.

\section{Results}
\label{results}

\begin{figure}
  \centering
  \includegraphics{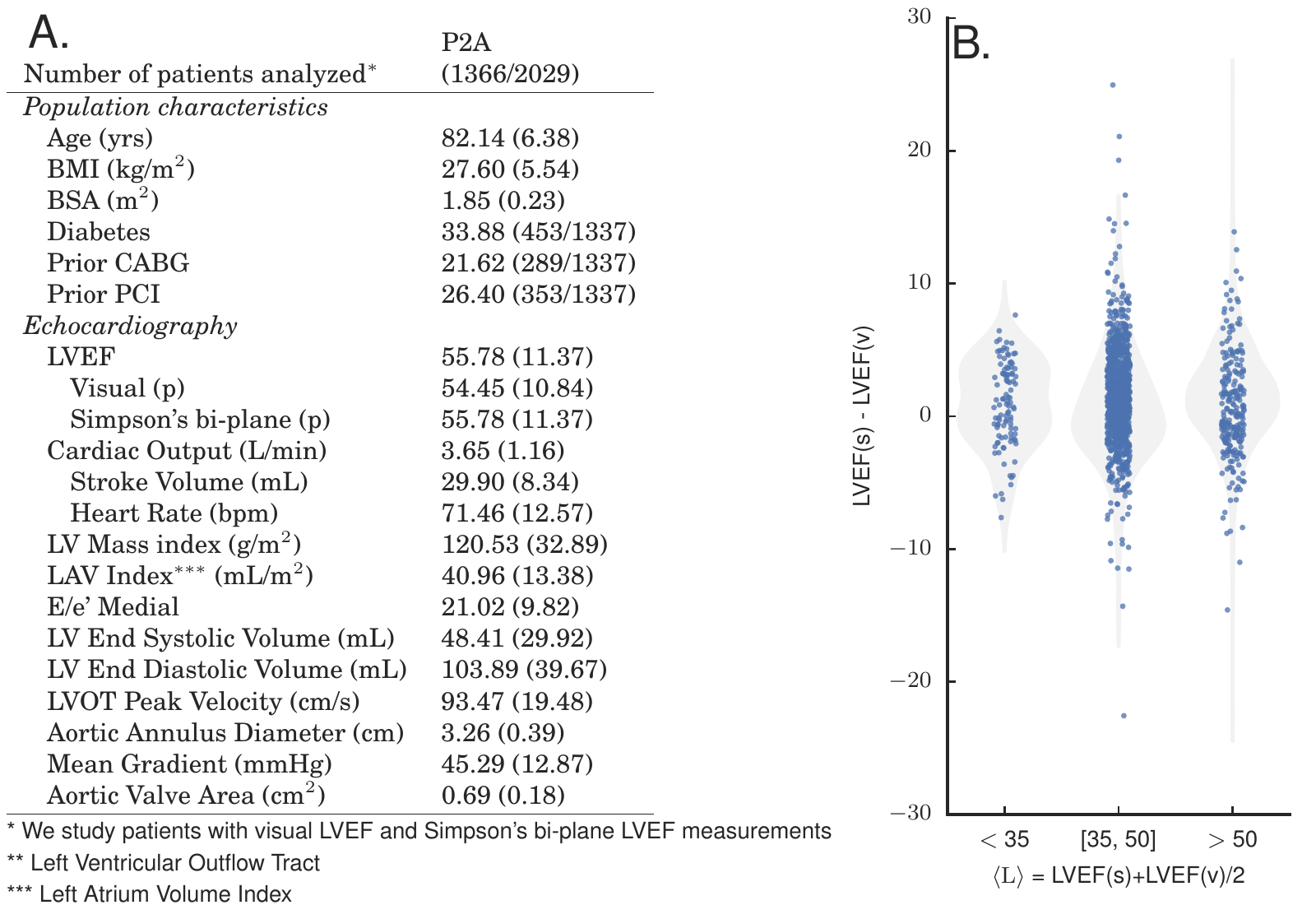}
  \caption{
    Patients with paired visual and Simpson's biplane estimates present the same clinical picture as PARTNER-IIA patients, and cardiologists took unbiased and agreeable LVEF measurements.
    (A) We report averages and standard deviation inside parentheses. Similar to PARTNER-IIA's ITT population, patients show signs of decreased heart function (low Cardiac output and small aortic diameter) and compensatory mechanisms (high LV mass).
    (B) For each of the 1366 patients with paired LVEF data, we plot the difference between Simpson's biplane method  and the cardiologist's visual estimate versus the average LVEF ($\mean{\mathrm{L}}$) between the two methods.
    Stratifying by an average LVEF, we find good agreement between Simpon’s biplane and visual estimates with an average difference of 0.93\% (s.d. 3.18\%) for $\mean{\mathrm{L}}$ $<$ 35\%, 1.53\% (s.d. 3.78\%) for $\mean{\mathrm{L}}$ between 35\% and 50\%, and 1.53\% (s.d. 3.78\%) for $\mean{\mathrm{L}}$ $>$ 50\%.
    Patients with paired data resemble the PARTNER-IIA population making this a realistic population to find in practice and show good agreement between Simpson’s biplane LVEF and the visual LVEF independent of average LVEF.
    \label{fig2.table1} }
\end{figure}

The PARTNER-IIA cohort contains $1366$ patients with paired visual estimates of LVEF and Simpson's biplane measured LVEF, and resembles the overall PARTNER-IIA cohort (Table~\ref{fig2.table1}A.).
Patients from this paired dataset have an average Simpson’s Ejection Fraction of 55.78\% (s.d. 11.37\%), moderate LV hypertrophy (mean LV mass index of $120.53$ g/$\mathrm{m}^{2}$ [s.d. $32.89$ g/$\mathrm{m}^{2}$]), and a reduced aortic valve area (0.69 c$\mathrm{m}^{2}$  [s.d. 0.18 c$\mathrm{m}^{2}$]) coupled with higher mean gradients (mean gradient $45.29$ mmHg [s.d. $12.87$ mmHg]).
The above results indicate these patients have baseline characteristics similar to the full cohort of PARTNER-IIA patients~\cite{leon2010transcatheter}.

Shifting focus to LVEF, we found no significant difference between Simpson's biplane LVEF and visual LVEF on average (Figure~\ref{fig2.table1}B). For patients with a visual LVEF under 35\%, the average difference between Simpson's LVEF and visual LVEF equals 0.93\% (s.d. 3.18\%).
Patients with visual LVEF estimates between 35\% and 50\% have an average difference between Simpson's LVEF and visual LVEF equal to 1.53\% (s.d. 3.78\%).
Patients with normal LVEF have an average difference between Simpson's LVEF and visual LVEF of 0.60\% (s.d. 4.04\%).
We find the maximum difference between Simpson's LVEF minus visual LVEF equals 24.96\%, and minimum difference between Simpson's LVEF minus visual LVEF equals -22.56\%. 

\begin{figure}
  \centering
  \includegraphics{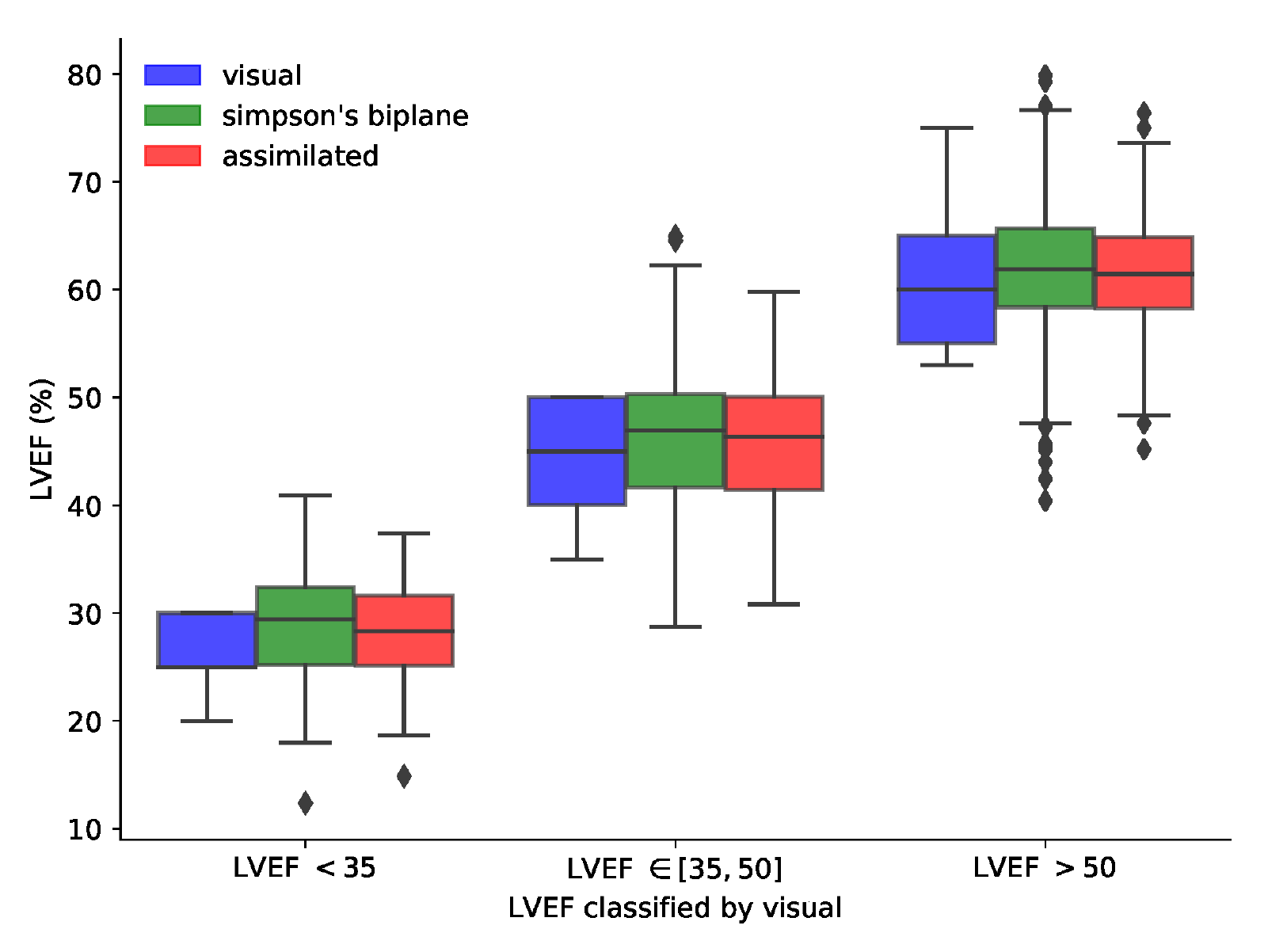}
  \caption{
    The assimilated LVEF achieves a population wide LVEF that lies closer to the more accurate Simpson's biplane method and with smaller variability than either visual or Simpson's biplane estimates of LVEF.
    We draw boxplots for each LVEF estimate method (visual, Simpson's biplane, and assimilated) stratified by visual LVEF below 35\%, LVEF between 35\% and 50\%, and LVEF above 50\%.
    Across strata, the assimilated median LVEF lies above the median visual LVEF and below Simpson’s biplane LVEF, and shrinks the interquartile range (length of boxes) compared to Simpson's biplane method.
    \label{fig3.bxPlots}
  }
\end{figure}

After fusing together the cardiologist's visual estimate with Simpson's biplane method, the assimilated LVEF shifted by +1.34\% (s.d. 1.97\%) on average compared to the visual LVEF and shifted by -0.65\% (s.d. 0.96\%) on average relative to Simpson's biplane method in patients with a visual LVEF $<$ 35\%, shifted by +0.82\% (s.d. 2.79\%) on average relative to visual LVEF and shifted by -0.40\% (s.d. 1.36\%) on average relative to Simpson's biplane method in patients with a visual LVEF estimate between 35\% and 50\%, and shifted by +0.87\%  (s.d. 2.51\%) on average relative to visual LVEF and by -0.42\% (s.d. 1.22\%) on average relative to Simpson's biplane method in patients with a visual LVEF above 50\% (Figure~\ref{fig3.bxPlots}).

We also found the assimilated LVEF reduces the population standard deviation (Figure~\ref{fig3.bxPlots}) relative to Simpson's biplane by 11.24\% in patients with visual LVEF $<$ 35\%, by 9.96\% in patients with visual LVEF between 35\% and 50\%, and by 13.21\% in patients with a visual LVEF above 50\%.

\begin{figure}
  \centering
  \includegraphics{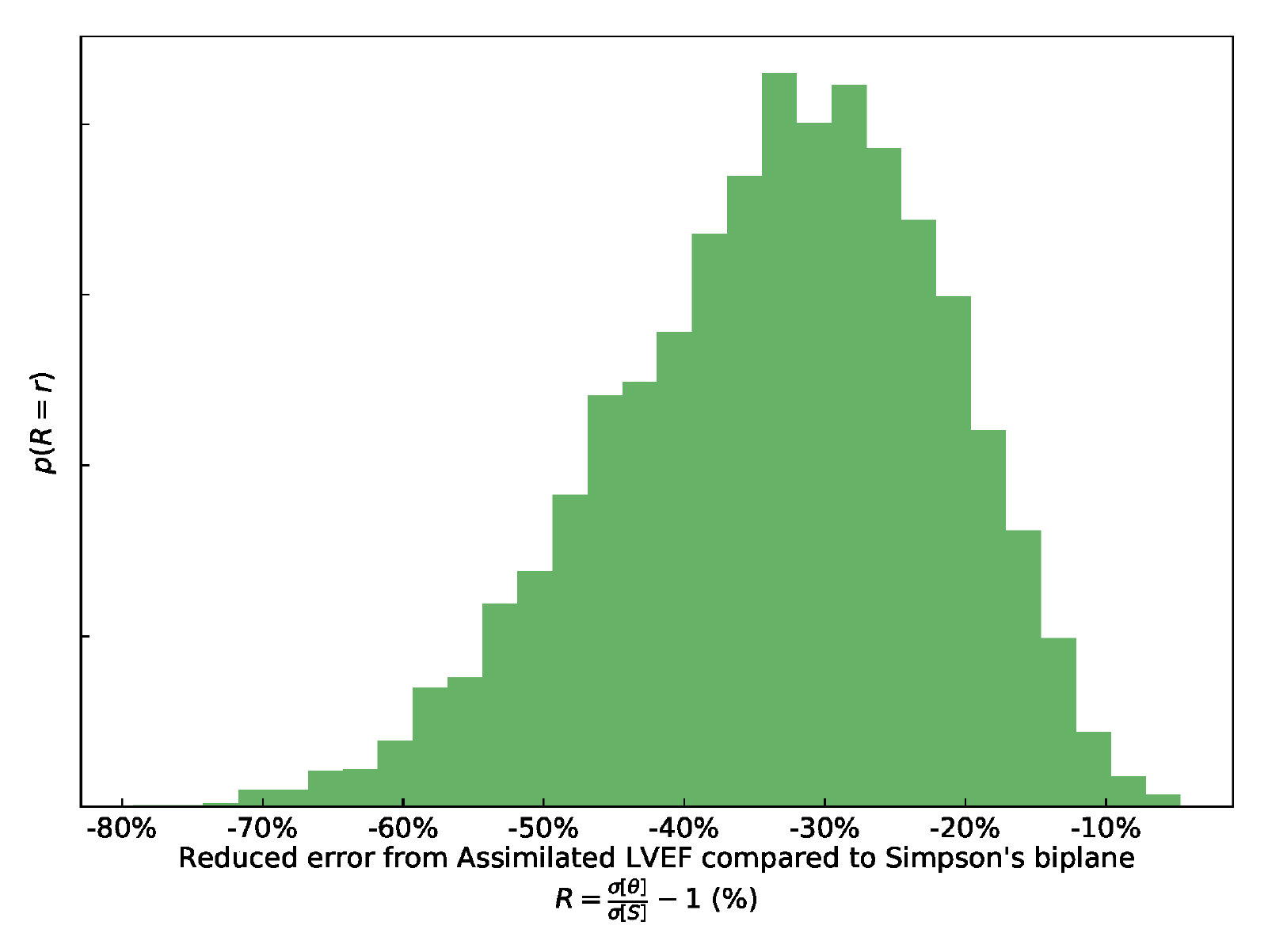}
  \caption{
    Compared to Simpson's biplane and considering our estimates of measurement error uncertain, our assimilated LVEF has an average -33.55\% (95\%CI[-58.69\%, -13.76\%]) relative reduction.
    This decrease in measurement error will lead to more clearly classifying patient with heart failure and more powerful statistical conclusions.
    \label{fig4.omega}}
\end{figure}

We more formally considered our uncertainty in visual and Simpson's measurement and found the assimilated LVEF reduced measurement error by -33.55\% (95\%CI [-58.69\%, -13.76\%]) on average compared to Simpson's biplane~(Figure.~\ref{fig4.omega}).
Our Bayesian model of Simpson's measurement error placed 95\% of the probability mass between 3.35\% and 17.72\%, with an average measurement error equal to 8.63\%.
Our visual measurement error's Bayesian model distributed 95\% of the probability mass between 8.00\% and 34.11\%, with an average measurement error equal to 17.68\%.
The most conservative circumstances still find the assimilated model reduces measurement error compared to Simpson's biplane.

\begin{figure}
  \centering
  \includegraphics{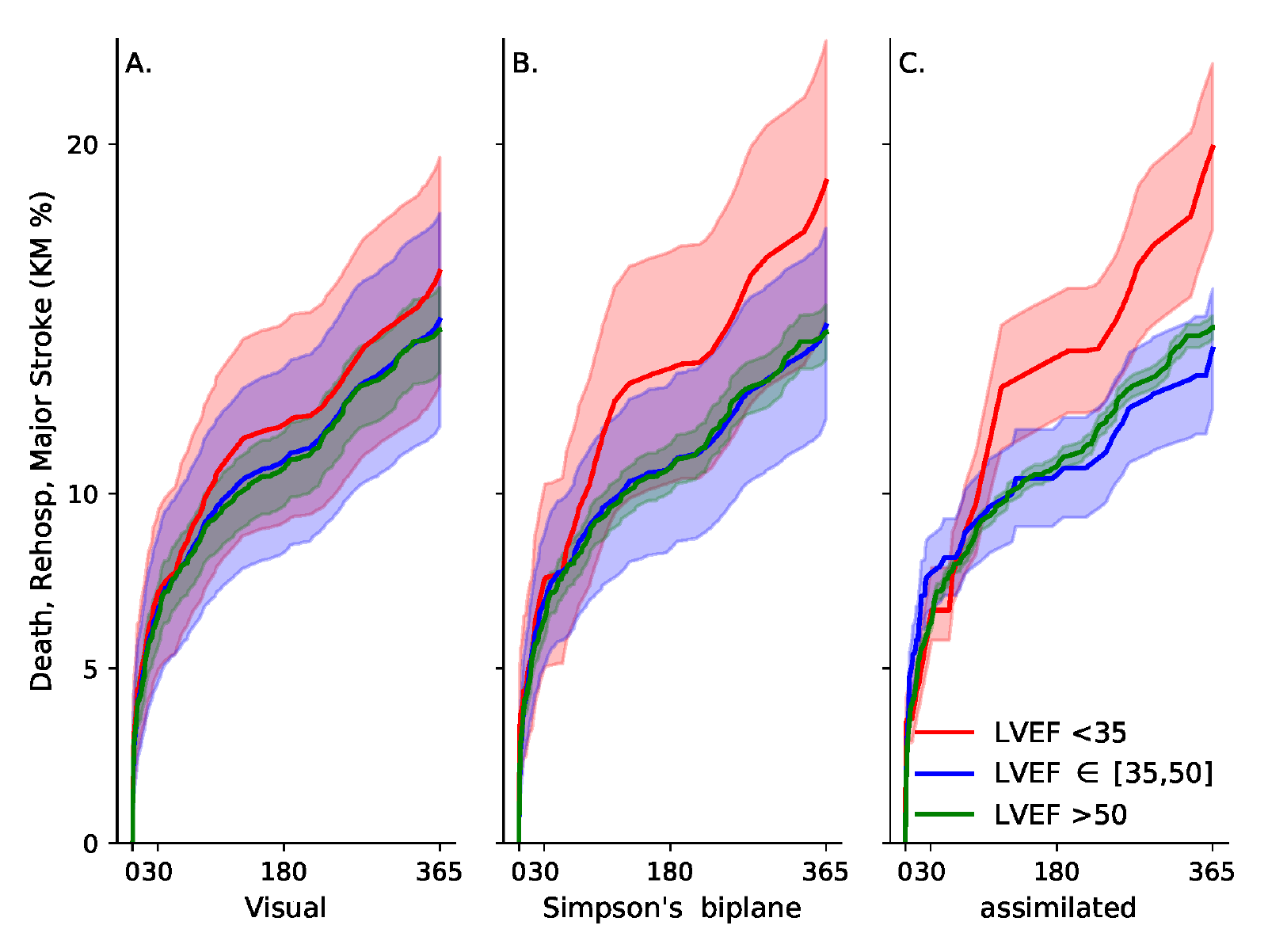}
  \caption{
    We quantify the impact of measurement error on the association between LVEF and 1 year death, stroke, and rehospitalization by (i) cutting the population into three groups based on visual LVEF, Simpson's biplane LVEF, and assimilated LVEF, (ii) estimating Kaplan-Meier event rates, and (iii) shading the event rate’s 95\% credible interval based on uncertainty in each patient's LVEF estimate.
    (A) Stratifying by visual LVEF shows minor differences between LVEF categories.
    (B) The average Kaplan-Meier curve's shape for patients stratified by Simpson’s biplane LVEF resembles the visual Kaplan-Meier curve, but has smaller credible intervals.
    LVEF measured by Simpson's biplane increased precision and reduced measurement error enough to draw statistical conclusions about patients with LVEF $<$ 35\%.
    (C) The assimilated LVEF shrinks credible intervals the most compared to visual or Simpson's biplane LVEF, and smaller credible intervals among strata translates to stronger conclusions about the association between LVEF and outcomes.   
    \label{fig5.KMS}
  }
\end{figure}

We explored the relationship of this novel assimilated LVEF with a composite of death, stroke, and rehospitalization at to 1 year, and found the strength of association between LVEF and outcome weakest for visually estimated LVEF, stronger for Simpson's biplane method, and strongest for using our assimilated method (Figure~\ref{fig5.KMS}).
The visually estimated LVEF measurements show the most variability with Kaplan-Meier's 95\% credible intervals overlapping throughout the full 365 days (Figure~\ref{fig5.KMS}A).
At one year, the event rate for patients with visually estimated LVEF $<$ 35\% was 16.35\% (95\% CI [13.08\%, 19.61\%]), for patients with visually estimated LVEF between 35\% and 50\% was 14.97\% (95\%CI [11.91\%, 18.02\%]), and for patients with visually estimated LVEF $>$ 50\% was 14.68\% (95\%CI [13.45\%, 15.89\%]).

We find smaller variability and more separation between patient groups classified by LVEF with Simpson's biplane method (Figure~\ref{fig5.KMS}B).
At one year, the event rate for patients with Simpson's biplane estimated LVEF $<$ 35\% equals 18.94\% (95\%CI [14.98\%, 22.96\%]), for patients with Simpson's biplane estimated LVEF between 35\% and 50\% equals 14.81\% (95\% CI [12.13\%, 17.60\%]), and for patients with Simpson's biplane estimated LVEF $>$ 50\% equals 14.62\% (95\%CI [13.83\%, 15.39\%]).

After we combine the cardiologist's visual LVEF with Simpson's LVEF measurement, we see the median failure rates moving closer together contrasted by a reduction in error bounds (Figure~\ref{fig5.KMS}C).
At 1-year, the event rate for patients with an assimilated LVEF $<$ 35\% equals 19.90\% (95\%CI [17.75\%, 22.30\%]), for patients with an assimilated LVEF between 35\% and 50\% equals 14.14\% (95\%CI [12.44\%, 15.85\%]), and for patients with an assimilated LVEF $>$ 50\% equal 14.75\% (95\%CI [14.42\%, 15.08\%]).

Compared to the visual LVEF measurement and Simpson's bi-plane measurement, the assimilated method pushed the LVEF patient group's average Kaplan-Meier estimate up and pulled the LVEF patient group's average Kaplan-Meier estimate down while at the same time reducing the estimate's overall variability (Figure~\ref{fig5.KMS}).

\begin{figure}
  \centering
  \includegraphics[scale=0.75]{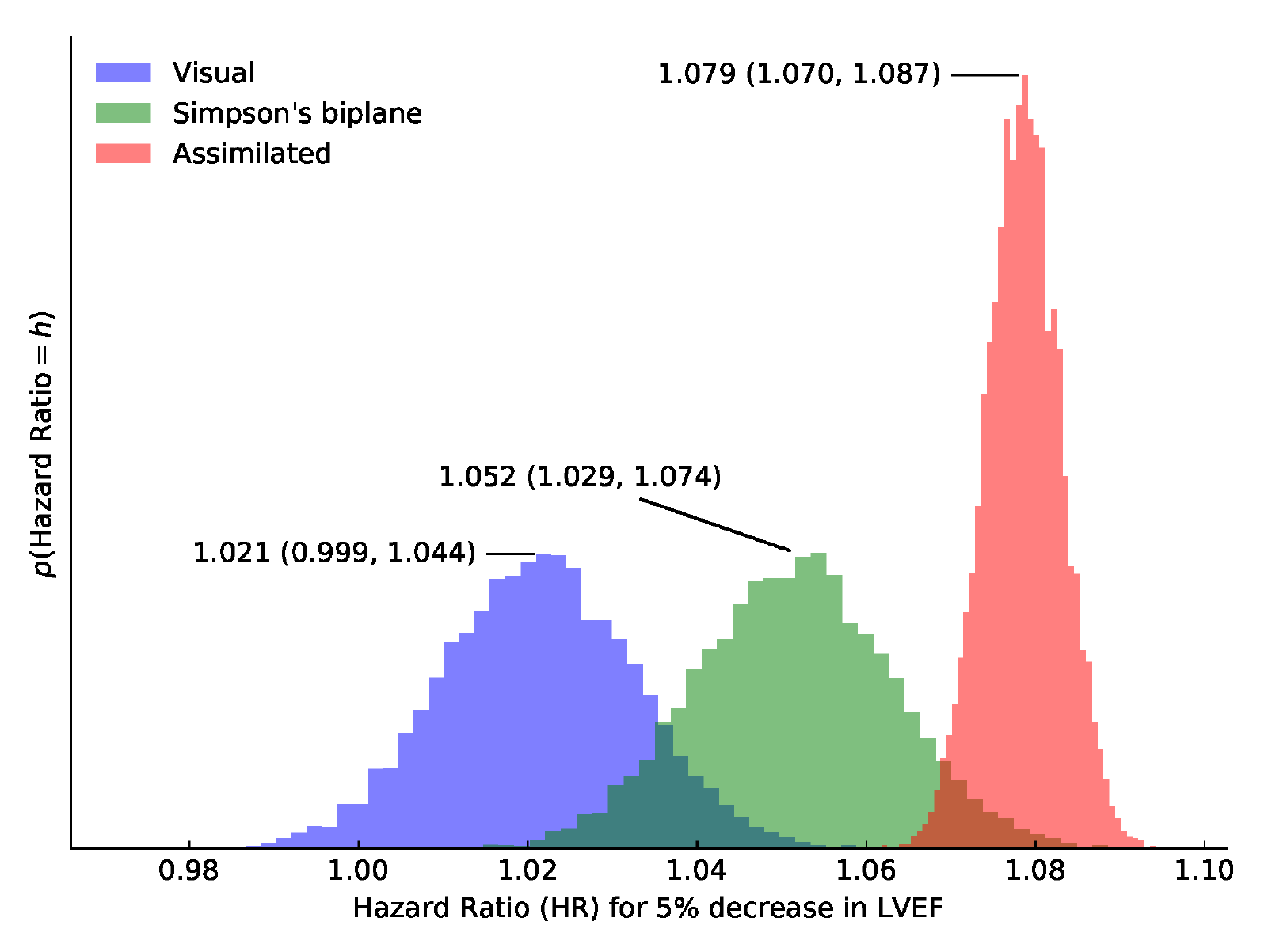}
  \caption{
    Estimated hazard ratios for a 5\% decrease in assimilated, Simpson's, and visual LVEF's impact on 1-year Death, stroke, and rehospitalization.
    We find the weakest association between visual LVEF and the 1-year endpoint (HR [95\%CI] = 1.021 [0.999, 1.044]), a stronger association between Simpson's LVEF and the 1-year endpoint (HR [95\%CI] = 1.052 [1.029, 1.074]), and the most consistent association between assimilated LVEF and the 1-year endpoint (HR [95\%CI] = 1.079 [1.070, 1.087]).
    The assimilated LVEF develops a more reproducible association with clinical endpoints. 
    \label{fig6.HRS}
  }
\end{figure}

The assimilated LVEF's hazard ratio relates decreasing LVEF and increasing hazards of 1-year death, stroke, and rehospitalization more confidently than Simpson's or visually estimated LVEF (Figure~\ref{fig6.HRS}).
We find a 5\% increase in assimilated LVEF corresponds to a 1.079 (95\%CI [1.070, 1.087]) hazard ratio, in Simpson's LVEF corresponds to a 1.052 (95\%CI [1.029, 1.074]) hazard ratio, and in visual LVEF corresponds to a 1.021 (95\%CI [0.999, 1.044]) hazard ratio.
The assimilated LVEF's smaller credible interval represents a more certain and reproducible relationship between LVEF and 1-year death, stroke, and rehospitalization.

\section{Discussion}
After combining the visually estimated LVEF, honed through years of training, with Simpson's method, derived from a simple LVEF tracing from apical views, we find this novel assimilated LVEF reduces measurement error, improves reproducibility, and better predicts hard clinical events at 1 year.

This study highlights the important role of reducing variability when measuring the most common clinical currency in heart disease - LVEF.
Combining the visual estimate and Simpson's biplane measurement blends the cardiologist's established expertise with accurate image analysis and outperforms simpler LVEF metrics.

In daily echocardiography practice, we unrealistically assume all physicians produce equal quality echo images.
Similar to assuming equal image quality, we do not adjust LVEF values for differing levels of experience / echocardiographic expertise.
Differing image quality and echocardiographic expertise could play a major role in visual LVEF and Simpson's biplane LVEF measurement accuracy.

This assimilated LVEF may prove useful in the presence of poor echocardiographic images, where the cardiologist's visual estimate can more accurately judge LVEF.
Upcoming work will investigate how qualitative measurements of image quality affect visual LVEF measurement and Simpson's LVEF measurement.  

We also aim to develop more sophisticated statistical models that further reduce measurement error.
Focusing on a single noisy variable allowed us to demonstrate the importance of noise, but we will expand our work to study multiple covariates under measurement error and how this affects their associations with outcomes.
As far as we know, small measurement errors do not alter associations between covariates, but we will explore whether many small measurement errors across many covariates can change relationships between variables.

We acknowledge several limitations for this study.
The cohort we studied was a subset of patients from the entire trial, though this subset did not differ in baseline characteristics from the overall cohort.
This analysis was not pre-specified and suffers from the same limitations of any retrospective analyses.
The estimates for LVEF variability depended on small samples from previous studies and needs more data to better estimate interobserver variability.

This work aspires toward a more robust way to study relationships between clinical measurements and outcomes, reduce measurement error, and shows the cardiologist's intuition (derived from years of experience and iterative self-correction) remains a valuable tool.

\section*{Acknowledgments}
The authors thank Karl~Lherisson and the Cardiovascular Research Foundation's Information Technology department for computational resources.

{\it Conflict of interest: None declared.}


\begin{thebibliography}{10}

\bibitem{hunt20092009}
Sharon~Ann Hunt, William~T Abraham, Marshall~H Chin, Arthur~M Feldman, Gary~S
  Francis, Theodore~G Ganiats, Mariell Jessup, Marvin~A Konstam, Donna~M
  Mancini, Keith Michl, et~al.
\newblock 2009 focused update incorporated into the acc/aha 2005 guidelines for
  the diagnosis and management of heart failure in adults: a report of the
  american college of cardiology foundation/american heart association task
  force on practice guidelines developed in collaboration with the
  international society for heart and lung transplantation.
\newblock {\em Journal of the American College of Cardiology}, 53(15):e1--e90,
  2009.

\bibitem{mcgrady2013n}
Michele McGrady, Christopher~M Reid, Louise Shiel, Rory Wolfe, Umberto Boffa,
  Danny Liew, Duncan~J Campbell, David Prior, and Henry Krum.
\newblock N-terminal b-type natriuretic peptide and the association with left
  ventricular diastolic function in a population at high risk of incident heart
  failure: results of the screening evaluation of the evolution of new-heart
  failure study (screen-hf).
\newblock {\em European journal of heart failure}, 15(5):573--580, 2013.

\bibitem{leon2010transcatheter}
Martin~B Leon, Craig~R Smith, Michael Mack, D~Craig Miller, Jeffrey~W Moses,
  Lars~G Svensson, E~Murat Tuzcu, John~G Webb, Gregory~P Fontana, Raj~R Makkar,
  et~al.
\newblock Transcatheter aortic-valve implantation for aortic stenosis in
  patients who cannot undergo surgery.
\newblock {\em New England Journal of Medicine}, 363(17):1597--1607, 2010.

\bibitem{leon2016transcatheter}
Martin~B Leon, Craig~R Smith, Michael~J Mack, Raj~R Makkar, Lars~G Svensson,
  Susheel~K Kodali, Vinod~H Thourani, E~Murat Tuzcu, D~Craig Miller, Howard~C
  Herrmann, et~al.
\newblock Transcatheter or surgical aortic-valve replacement in
  intermediate-risk patients.
\newblock {\em N Engl J Med}, 2016(374):1609--1620, 2016.

\bibitem{ho1993epidemiology}
Kalon~KL Ho, Joan~L Pinsky, William~B Kannel, and Daniel Levy.
\newblock The epidemiology of heart failure: the framingham study.
\newblock {\em Journal of the American College of Cardiology}, 22(4):A6--A13,
  1993.

\bibitem{ho1993survival}
Kalon~KL Ho, Keaven~M Anderson, William~B Kannel, William Grossman, and Daniel
  Levy.
\newblock Survival after the onset of congestive heart failure in framingham
  heart study subjects.
\newblock {\em Circulation}, 88(1):107--115, 1993.

\bibitem{consensus1987effects}
CONSENSUS Trial~Study Group et~al.
\newblock Effects of enalapril on mortality in severe congestive heart failure:
  results of the cooperative north scandinavian enalapril survival study
  (consensus).
\newblock {\em N Engl j Med}, 316:1429--1435, 1987.

\bibitem{cleland2005effect}
John~GF Cleland, Jean-Claude Daubert, Erland Erdmann, Nick Freemantle, Daniel
  Gras, Lukas Kappenberger, and Luigi Tavazzi.
\newblock The effect of cardiac resynchronization on morbidity and mortality in
  heart failure.
\newblock {\em New England Journal of Medicine}, 352(15):1539--1549, 2005.

\bibitem{witte2005effect}
Klaus~KA Witte, Nikolay~P Nikitin, Anita~C Parker, Stephan von Haehling,
  Hans-Dieter Volk, Stefan~D Anker, Andrew~L Clark, and John~GF Cleland.
\newblock The effect of micronutrient supplementation on quality-of-life and
  left ventricular function in elderly patients with chronic heart failure.
\newblock {\em European heart journal}, 26(21):2238--2244, 2005.

\bibitem{jaarsma2000self}
Tiny Jaarsma, R\_ Halfens, F~Tan, H~Huijer Abu-Saad, K~Dracup, and J~Diederiks.
\newblock Self-care and quality of life in patients with advanced heart
  failure: the effect of a supportive educational intervention.
\newblock {\em Heart \& Lung: The Journal of Acute and Critical Care},
  29(5):319--330, 2000.

\bibitem{demers2001reliability}
Catherine Demers, Robert~S McKelvie, Abdissa Negassa, Salim Yusuf, RESOLVD
  Pilot~Study Investigators, et~al.
\newblock Reliability, validity, and responsiveness of the six-minute walk test
  in patients with heart failure.
\newblock {\em American heart journal}, 142(4):698--703, 2001.

\bibitem{gavazzi1997value}
Antonello Gavazzi, Carlo Berzuini, Carlo Campana, Corinna Inserra, Marina
  Ponzetta, Roberta Sebastiani, Stefano Ghio, and Franco Recusani.
\newblock Value of right ventricular ejection fraction in predicting short-term
  prognosis of patients with severe chronic heart failure.
\newblock {\em The Journal of heart and lung transplantation: the official
  publication of the International Society for Heart Transplantation},
  16(7):774--785, 1997.

\bibitem{christenson1995effect}
Jan~T Christenson, J~Maurice, F~Simonet, A~Bloch, PC~Fournet, V~Velebit, and
  M~Schmuziger.
\newblock Effect of low left ventricular ejection fractions on the outcome of
  primary coronary by-pass grafting in end-stage coronary artery disease.
\newblock {\em The Journal of cardiovascular surgery}, 36(1):45--51, 1995.

\bibitem{gudmundsson2005visually}
Petri Gudmundsson, Erik Rydberg, Reidar Winter, and Ronnie Willenheimer.
\newblock Visually estimated left ventricular ejection fraction by
  echocardiography is closely correlated with formal quantitative methods.
\newblock {\em International journal of cardiology}, 101(2):209--212, 2005.

\bibitem{bellenger2000comparison}
NG~Bellenger, MI~Burgess, SG~Ray, A~Lahiri, AJS Coats, JGF Cleland, and
  DJ~Pennell.
\newblock Comparison of left ventricular ejection fraction and volumes in heart
  failure by echocardiography, radionuclide ventriculography and cardiovascular
  magnetic resonance. are they interchangeable?
\newblock {\em European heart journal}, 21(16):1387--1396, 2000.

\bibitem{otterstad1997accuracy}
JE~Otterstad, G~Froeland, M~St~John Sutton, and I~Holme.
\newblock Accuracy and reproducibility of biplane two-dimensional
  echocardiographic measurements of left ventricular dimensions and function.
\newblock {\em European heart journal}, 18(3):507--513, 1997.

\bibitem{van1996comparison}
Niels van Royen, Carl~C Jaffe, Harlan~M Krumholz, Kevin~M Johnson, Patrick~J
  Lynch, Donna Natale, Patricia Atkinson, Paul Demon, J~Th Frans, et~al.
\newblock Comparison and reproducibility of visual echocardiographic and
  quantitative radionyclide left ventricular ejection fractions.
\newblock {\em American Journal of Cardiology}, 77(10):843--850, 1996.

\bibitem{chuang2000importance}
Michael~L Chuang, Mark~G Hibberd, Carol~J Salton, Raymond~A Beaudin, Marilyn~F
  Riley, Robert~A Parker, Pamela~S Douglas, and Warren~J Manning.
\newblock Importance of imaging method over imaging modality in noninvasive
  determination of left ventricular volumes and ejection fraction: assessment
  by two-and three-dimensional echocardiography and magnetic resonance imaging.
\newblock {\em Journal of the American College of Cardiology}, 35(2):477--484,
  2000.

\end{thebibliography}

\end{document}